\documentclass[%
 reprint,
 amsmath,amssymb,
 aps,
showkeys]{revtex4-2}

\usepackage{graphicx}% Include figure files
\usepackage{dcolumn}% Align table columns on decimal point
\usepackage{bm}% bold math
\usepackage{array,supertabular,hhline,enumitem,hyperref}
\usepackage{placeins}
\usepackage{float}
\usepackage{multirow}
\usepackage{makecell}
\usepackage{titlesec}

\titlespacing*{\section}{0pt}{2ex}{2ex}
\begin{document}

%\preprint{APS/123-QED}

\title{Formalizing Poisson-Boltzmann Theory for Field-Tunable Nanofluidic Devices}% Force line breaks with \\

\author{Zhongyuan Zhao}
\author{Chudi Qi}
\author{Yuheng Li}
\author{Shoushan Fan}
\author{Qunqing Li}
\author{Yang Wei}
 \email{weiyang@tsinghua.edu.cn}
 \affiliation{Department of Physics and
 	Tsinghua-Foxconn Nanotechnology Research Center, State Key Laboratory of Low-Dimensional Quantum Physics, Tsinghua University, Beijing 100084, China.}


\begin{abstract}
Nanofluidic devices support unconventional ion transport appealing to energy and information technologies, thanks to the susceptibility of confined electric double layers (EDLs) to various external physical fields. Although experimental studies advance rapidly, the rationalization of field-tunable nanofluidic transport has not reached a formalized and unified level. Here we reformulate the Poisson-Boltzmann theory and reveal distinct EDL regimes on the parameter space. Based on the regime classification, we establish a formal framework for the tunable nanofluidic transport and discuss the electrostatic modulation (ionic transistor) extensively. The framework reproduces the observed conductivity-concentration scaling behaviors, rationalizes the ionic transistors with reconfigurable polarities, and predicts one fundamental thermodynamic limit for electrostatic modulation (60 mV/dec). Being accurate, generalizable and extensible, this framework can account for a wide range of ion transport in confined spaces.
%\begin{description}
%\end{description}
\end{abstract}

\keywords{Nanofluidics, iontronics, ion transport, electric double layers, Poisson-Boltzmann theory}%[showkeys]%Use showkeys class option if keyword
\maketitle

%\tableofcontents

\section{Introduction}
Nanofluidic devices facilitate ion transport within nanoconfined spaces, where the ionic motions are strongly regulated by the electric double layer (EDL) effect \cite{006,007,008,009}. The confined EDLs induce novel transport properties, including surface sensitivity, ionic selectivity , and ionic rectification \cite{011,016,017,018,019,020}.  Thus being superior to bulk counterparts, nanofluidic devices have become increasingly significant in energy, biomedical and information technologies. More fundamentally, EDL effect naturally couples to hydrodynamic, mechanical, photochemical, and even quantum effects \cite{012,021,022,023,024,025}. Such physical couplings provide abundant opportunities for external field modulation, realizing biomimetic ionic transistors, amplified osmotic energy converters, and multi-modal ionic sensors \cite{026,027,028,029,030,031}. However, despite the surge of experimental breakthroughs \cite{010,011,012,013,014,015}, effective theories for field-tunable ionic transport remain insufficient \cite{008,009}. 

On the continuum and mean-field level, the Poisson-Boltzmann (PB) theory describes the ionic population in EDLs and resolves the steady-state transports  \cite{032,033,037,038,039,040}. However, the complexity of nanoconfinement hinders a global closed-form analytical solution to the PB equation. Pioneering studies have analytically treated the PB theory under certain approximations and rationalized the nanofluidic transports in specific ion channels \cite{010,014,034,035}. Yet the external field modulation remains unaddressed. While more recent modelings based on finite element methods can give accurate PB solutions and incorporate external fields \cite{017,021,036}, they're short in generally mapping specific numerical results to realistic devices and providing clear physical insights. Since both analytical and numerical methods have limitations, a unified physical picture for the confined-EDL problem over the entire parameter space is elusive, and a formal framework that comprehensively describes the nanofluidic transport under external physical fields is still lacking.

In this work, we firstly unite analytical and numerical methods, and present a formal reformulation of the PB theory. Mapping realistic EDL parameters to confinement factor $\gamma$ and relative field strength $\chi$, the nondimensionalized PB equation is numerically solved over the $(\gamma,\chi)$ parameter space. A set of regime indicators are defined, by which EDL regimes with distinct characteristics are rigorously classified. The separation of linear-response, EDL-overlap and surface accumulation regimes, is revealed to be the physical origin of nanofluidic transport diversities. We then establish a formal framework for field-tuned transport. By relating external tuning parameter $\mathcal{F}$ to ionic conductivity $G$ through $(\gamma,\chi)$, the tunable transport behaviors $G(\mathcal{F})$ can be precisely described, as verified by the reproductions of $G - n_0$ scaling behaviors. Electrostatic modulations by a gate voltage $\Phi_\mathrm{g}$ are extensively investigated, demonstrating the possibility of reconfigurable ionic transistors via surface modification. Significantly, we uncover one fundamental thermodynamic limit for the electrostatic modulation efficiency, namely the subthreshold swing (SS), which is 60 mV/dec  at room temperature. This framework can be generalized by introducing various physical fields $\mathcal{F}$ and can be extended with finer terms for $G$. These advances deepen the understanding of field-tunable ionic transport and provide direct guidance for the optimization of nanofluidic devices.
\begin{figure*}[htbp]
	\includegraphics[width=0.8\textwidth]{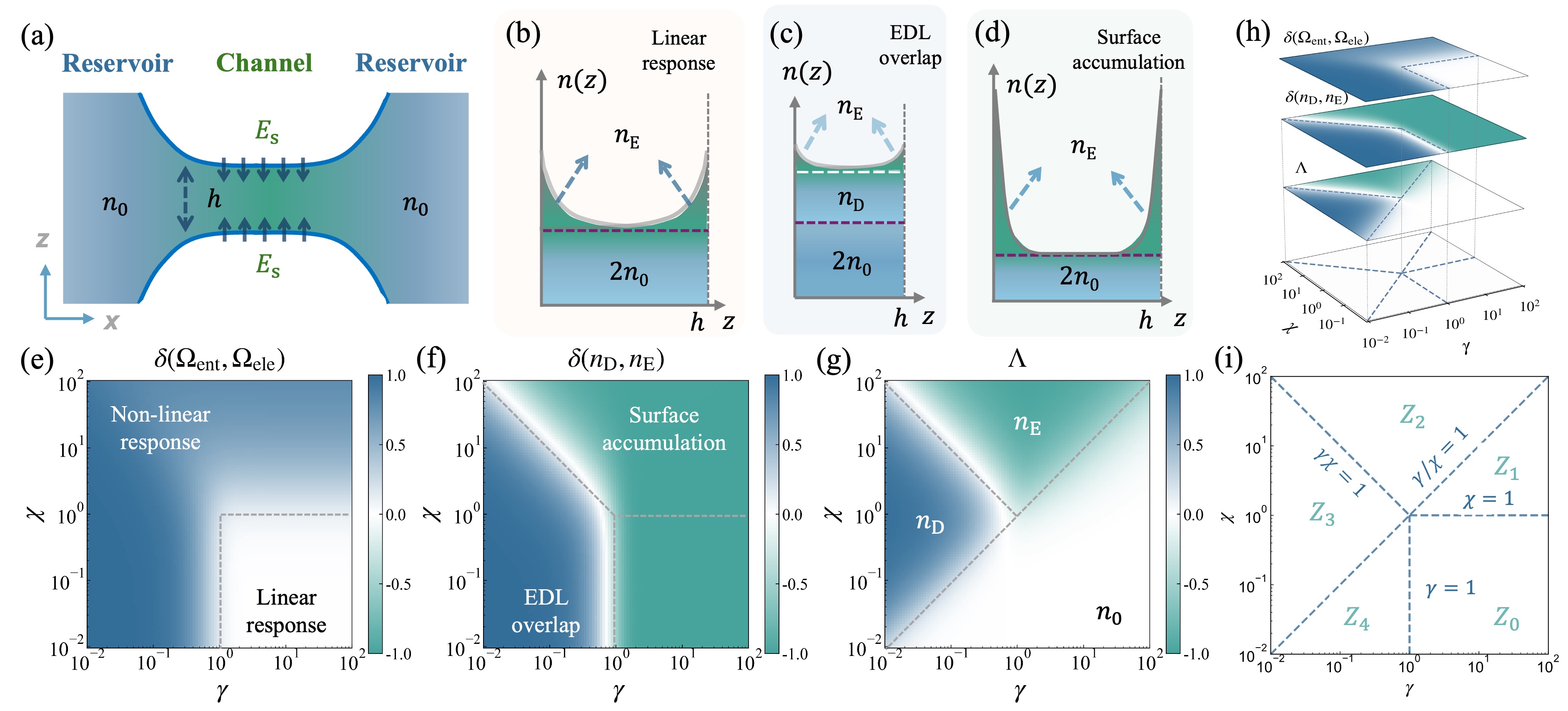}% Here is how to import EPS art
	\caption{(a) The configuration of the nanochannel model. (b-d) Schematic ion concentration profiles in the
		linear-response regime (b), surface-accumulation regime (c) and EDL-overlap regime (d). (e)-(g) Mappings of the "order parameters". (h) Stacking of the three mappings. (i) The regime diagram for PB theory.}
	\label{fig:Figure 1}
\end{figure*}
\section{Model}
A general reservoir-channel-reservoir model is considered, as depicted in Fig.~\ref{fig:Figure 1}(a). The reservoirs have an ionic concentration  $n_0$, while the channel is confined to $h$ and is subjected to surface electric fields  $E_\mathrm{s}$ at both channel walls (see  Supplementary Materials, SM, section 1.1). Due to  $E_\mathrm{s}$, counter-ions are drawn from the reservoir into the channel while co-ions are repelled, forming
an electrically nonneutral region, known as the electric double layer (EDL).  This redistribution is associated
with electric potential variation  $\varphi (z)$ relative to the reservoir. At equilibrium, ions follow
Boltzmann distribution  $n_{\pm }(z)=n_0\exp (-z_{\pm }\mathit{e\varphi }/k_BT)$. By incorporating Poisson equation 
${\nabla}^2\varphi =-e(n_{+}z_{+}+n_{-}z_{-})/\epsilon $, the Poisson-Boltzmann (PB) equation is derived as:
\begin{equation}
	{\nabla}^2\Phi =\kappa ^2\sinh \Phi ,
\end{equation}
for a monovalent symmetric ($z_{\pm}=\pm1$) electrolyte. $\Phi =\mathit{e\varphi}/k_\mathrm{B}T$ is the reduced electric potential and  $\kappa^{-1}=l_\mathrm{D}=\sqrt{\mathit{\epsilon k}_\mathrm{B}T/2e^2n_0}$ is the
Debye length. Model parameters  $n_0$,  $h$ and $E_\mathrm{s}$ can be linked to length scales:
\begin{equation}
	l_\mathrm{D}=\sqrt{\mathit{\epsilon k}_\mathrm{B}T/2e^2n_0},\quad h=h,\quad l_{\mathrm{GC}}=2k_\mathrm{B}T/eE_\mathrm{s}.
\end{equation}
The physical implications of Debye length  $l_\mathrm{D}$  and Gouy-Chapman length $l_{\mathrm{GC}}$  are unraveled in the next section. 

We primarily focus on ionic population  $n\left(\mathbf{r}\right)=n_{+}+n_{-}$ \ and free energy  $\Omega$, as the former governs ionic conductivity $G$ and the latter is encoded with response characteristics to external fields in its derivatives. In the simplest nano-slit configuration ($\varphi$ and  $n$ vary only in the $z$ direction,  $z{\in}\left[0,h\right], |\varphi_\mathrm{s}^{'}|=E_\mathrm{s}$), the averaged total
ionic concentration $n_\mathrm{t}$ is (detailed in SM, section 2):
\begin{eqnarray}
	&n_\mathrm{t}&=\frac 1 h\int _0^h\left(n_{+}+n_{-}\right)\mathrm{d}z=2n_0+n_\mathrm{E}+n_\mathrm{D},
	\label{eq_total_density}\\
	&n_\mathrm{E}&=\!\frac 1{hk_\mathrm{B}T}\!\int _0^h\!\!\frac{\epsilon } 2\!\left(\!\frac{\mathrm{d}\varphi
		}{\mathrm{d}z}\right)^2\!\!\mathrm{d}z,\\
		&n_\mathrm{D}&=2n_0\!\left(\cosh \Phi _\mathrm{D}-1\right).
		\label{eq_n_E_n_D}
\end{eqnarray}
$n_\mathrm{E}$ is associated with spatial variations of $\Phi $, capturing the non-uniform accumulation near the surface (Debye screening). $n_\mathrm{D}$ captures the uniform accumulation across the
channel, which marks the confinement-induced EDL-overlap and relates to a non-zero mid-channel Donnan potential  $\Phi _\mathrm{D}$ \cite{040}. The free energy consists of the entropic  $\Omega
_{\mathrm{ent}}$ and the electrostatic  $\Omega _{\mathrm{ele}}$ \cite{038}:
\begin{eqnarray}
\Omega _{\mathrm{ent}}&=&k_\mathrm{B}T\sum _{\alpha =\pm }\int \left(n_{\alpha }\ln \frac{n_{\alpha }}{n_0}-n_{\alpha
		}+n_0\right)\mathrm{d}z,\\
		\Omega _{\mathrm{ele}}&=&\int \frac{\epsilon } 2\left({\nabla}\varphi
		\right)^2\mathrm{d}z.
		\label{eq_free_energies}
\end{eqnarray}
Given model parameters, $\Phi (z)$, $n_\mathrm{D/E}$ and $\Omega _{\mathrm{ent/ele}}$ can be solved from PB equation, and the nanofluidic transport properties can be resolved. This is a mean-field and continuum treatment. We ignore ion-ion interactions or correlations, possible chemical reactions and hydration-shell complexities, and describe only the redistribution of point-like ions in a solution environment with dielectric constant $\epsilon$. Being simple though, this treatment  is believed to be applicable down to 1-2 nm \cite{006,009,053}, and should cover a wide range of nanofluidic devices. Starting with the minimal model, we shall capture the key ingredients of a field-tunable nanofluidic device, incrementally incorporate more sophisticated effects, and give guiding principles for experimental optimization.
\section{The Structure of Poisson-Boltzmann theory}
\begin{table*}
	\caption{Summary of EDL regimes and their respective characteristics.}
	\label{tab:Table_1}
	\centering
	\renewcommand{\arraystretch}{1.05}
	\setlength{\tabcolsep}{6.0pt}
	\begin{tabular}{|c|c|c|c|c|c|}
		\hline
		Zone & $Z_3$ & $Z_4$ & $Z_0$ & $Z_1$ & $Z_2$ \\
		\hline
		EDL feature &
		\multicolumn{2}{c|}{\begin{tabular}{c}
				EDLs overlap \\
				$(\gamma < 1, \gamma\chi < 1)$
		\end{tabular}} &
		\begin{tabular}{c}
			Linear response \\
			$(\gamma > 1,\ \chi < 1)$
		\end{tabular} &
		\multicolumn{2}{c|}{\begin{tabular}{c}
				Surface accumulation \\
				$(\chi > 1,\ \gamma\chi > 1)$
		\end{tabular}}
		\\
		\hline
		Characteristic length &
		\multicolumn{2}{c|}{$h$} &
		$l_\mathrm{D}=\sqrt{\varepsilon k_\mathrm{B} T / 2 e^2 n_0}$ &
		\multicolumn{2}{c|}{$l_\mathrm{GC}=2k_\mathrm{B} T/eE_\mathrm{s}$}
		\\
		\hline
		Concentration dominance &{
			\begin{tabular}{c}
				$n_\mathrm{D}$-dominant \\
				$(\gamma\chi < 1,\ \chi/\gamma > 1)$
		\end{tabular}}&
		\multicolumn{3}{c|}{
			\begin{tabular}{c}$n_0$-dominant \\
				$(\chi/\gamma < 1)$\end{tabular}}&
		{\begin{tabular}{c}
				$n_\mathrm{E}$-dominant \\
				$(\gamma\chi > 1,\ \chi/\gamma > 1)$
		\end{tabular}}
		\\
		\hline
		Related effects &
		\multicolumn{2}{c|}{\begin{tabular}{c}
				Charge selectivity \cite{017} \\
				Ionic rectification \cite{019}\\
				Transistor behavior \cite{027}
		\end{tabular}} &
		Trivially bulk like &
		\multicolumn{2}{c|}{\begin{tabular}{c}
				Electro-osmosis \cite{010,020}\\
				Ion-ion correlations \cite{032}
		\end{tabular}}
		\\
		\hline
	\end{tabular}
\end{table*}
The structure of the PB theory is investigated, which is a critical step for a formal description of nanofluidic transport. Nondimensionalization is introduced via $\gamma$ and $\chi$:
\begin{equation}
	\gamma =\frac h{4l_\mathrm{D}}{\propto}h\sqrt{n_0},\quad\chi
	=\frac{l_\mathrm{D}}{l_{\mathrm{GC}}}{\propto}\frac{E_\mathrm{s}}{\sqrt{n_0}},
\end{equation}
(a prefactor 1/4 is introduced to  $\gamma$, to gain concise mathematical forms, as we will show later) and the 1D PB equation is recast with respect
to $\overline z=z/l_\mathrm{D}$ in a dimensionless form:
\begin{equation}
	\frac{\mathrm{d}^2\Phi }{\mathrm{d}\overline{z}^2}=\sinh \Phi ,\quad\overline{z}{\in}\left[0,4\gamma \right],\quad|\Phi
	_\mathrm{s}^{'}|=2\chi ,
\end{equation}
in which $\Phi _\mathrm{s}^{'}$ is the derivative of  $\Phi$ to $\overline{z}$ at the channel walls.  $\gamma$ characterizes the extent of spatial confinement and  $\chi $ is the relative field strength.  For confined EDLs, intricate competitions among entropic, electrostatic and confinement effects hinder a global closed-form PB solution. Here, customized numerical methods (SM, section 4) yield accurate PB solutions over the entire parameter space ($\gamma$ and $\chi$). We define a set of regime indicators:
\begin{eqnarray}
	&\delta \left(\Omega _{\mathrm{ent}}, \Omega _{\mathrm{ele}}\right)&=\frac{\Omega _{\mathrm{ent}}-\Omega
			_{\mathrm{ele}}}{\Omega _{\mathrm{ent}}+\Omega _{\mathrm{ele}}},\\
			&\delta\left(n_\mathrm{D},n_\mathrm{E}\right)&=\frac{n_\mathrm{D}-n_\mathrm{E}}{n_D+n_\mathrm{E}},\\
			&\Lambda&=\frac{n_\mathrm{D}-n_\mathrm{E}}{n_\mathrm{D}+n_\mathrm{E}+2n_0}.
\end{eqnarray}

EDL regimes with distinct features can be efficiently classified according to the mappings of the three indices shown in Figs.~\ref{fig:Figure 1}(e)-\ref{fig:Figure 1}(g). The first is the linear-response regime, shown by the white-colored region in Fig.~\ref{fig:Figure 1}(e) ($\delta \left(\Omega _{\mathrm{ent}},\Omega _{\mathrm{ele}}\right)\rightarrow 0$ at approximately $\gamma
>1$,  $\chi<1$, or equivalently under weak confinement  $h>l_\mathrm{D}$ and weak field $l_{\mathrm{GC}}>l_\mathrm{D}$). In this regime, the PB equation can be linearized to the Debye-Hückel form \cite{041,042}:  $\ddot{\Phi }=\kappa ^2\Phi $, resembling linear harmonic oscillator, and naturally yielding the equipartition between  $\Omega _{\mathrm{ent}}$ and $\Omega _{\mathrm{ele}}$. This balance gives an ionic accumulation diminishing toward channel interior over the Debye length $l_\mathrm{D}$, which constitutes the best-known picture of EDL and is illustrated in terms of $n\left(z\right)=n_{+}+n_{-}$ by Fig.~\ref{fig:Figure 1}(b). 

The deviation from equipartition in Fig.~\ref{fig:Figure 1}(e) (blue color,  $\Omega_{\mathrm{ent}}{\gg}\Omega _{\mathrm{ele}}$) signs non-linear regimes, including the EDL-overlap regime and the surface-accumulation regime. In the EDL-overlap regime (approximately  $\gamma <1$ and  $\mathit{\gamma \chi }<1$, or equivalently  $l_\mathrm{D}>h$ \ and  $l_{\mathrm{GC}}>h$), strong spatial confinement prevents the channel interior from being fully screened, raising the electric potential by  $\Phi _\mathrm{D}$. As shown by the blue region in Fig.~\ref{fig:Figure 1}(f), $n_\mathrm{D}$ prevails over  $n_\mathrm{E}$, indicating that the counter-ions are almost uniformly enriched across the channel (Fig.~\ref{fig:Figure 1}(d)). This EDL-overlap behavior is exclusive to nanoconfinement effect ($h<l_\mathrm{D}$) and gives rise to ionic selectivity, rectification and transistor behaviors \cite{017,019,027}. In the surface-accumulation regime (approximately  $\chi >1$ \ and $\mathrm{\gamma \chi }>1$, or equivalently  $l_\mathrm{D}>l_{\mathrm{GC}}$ and  $h>l_{\mathrm{GC}}$), the strong surface field $E_\mathrm{s}$ pulls counter-ions tightly to the channel surface, forming a highly localized ion accumulation  $n_\mathrm{E}{\gg}n_\mathrm{D}$, indicated by the green-colored region (excluding the linear-response regime) in Fig.~\ref{fig:Figure 1}(f) and illustrated in Fig.~\ref{fig:Figure 1}(c). Counter-ions are localized within $l_{\mathrm{GC}}=2k_\mathrm{B}T/eE_\mathrm{s}$, which is the effective force range of a charged plane under thermal agitation (SM, section 3.2). Such a tight ionic population near the surface yields large electric potential gradients, which couples to the hydrodynamic effects \cite{010,020} and may induce  ion-ion correlations and even quantum effects \cite{032,025}. 

In terms of ion numbers, $\Lambda $ signifies that, when $\chi /\gamma <1$, $\Lambda \rightarrow 0$,  $n_0$ outnumbers both $n_\mathrm{D}$ and  $n_\mathrm{E}$ , resulting in a bulk-like channel (white region in Fig.~\ref{fig:Figure 1}(g)). While for  $\chi /\gamma >1$, accumulation in EDL prevails either by  $n_\mathrm{D}$ (blue region,  $\Lambda \rightarrow 1$) or  $n_\mathrm{E}$ (green region, $\Lambda \rightarrow -1$).

The three mappings in Fig.~\ref{fig:Figure 1} are stacked in Fig.~\ref{fig:Figure 1}(h). Though the crossovers between different regimes are smooth and continuous, we can still intuitively plot the four dividing lines ($\gamma =1$, $\chi =1$, $\mathrm{\gamma \chi }=1$, and $\chi /\gamma =1$), and yield a unifying regime diagram in Fig.~\ref{fig:Figure 1}(i). The four dividing lines partition the parameter plane into five zones ($Z_0$ to  $Z_4$) and are summarized in Table.~\ref{tab:Table_1} . Inspired by regime-classification, regime-specific heuristic forms are derived for surface potential  $\Phi _\mathrm{s}$ and for relative excess ion concentration  $\Gamma =(n_\mathrm{D}+n_\mathrm{E})/2n_0$: 
\begin{eqnarray}
	\Phi _\mathrm{s}&=&\left\{
	\begin{matrix}
		\frac{2\chi }{\tanh 2\gamma },\quad Z_0;\quad\\
		2\ln \left(2\chi\right), \quad Z_1{\cup}Z_2\\ 
		\mathrm{arcsinh}\left(\frac{\chi }{\gamma}\right), \quad Z_3{\cup}Z_4
	\end{matrix}\right.\\
		\Gamma &=&\left\{
		\begin{matrix}\frac 1{\gamma }\left(\sqrt{1+\chi^2}-1\right), \quad Z_0{\cup}Z_1{\cup}Z_2\\
			\sqrt{1+(\frac{\chi }{\gamma})^2}-1, \quad Z_3{\cup}Z_4
		\end{matrix}\right.
	\label{eq_Gamma_expression}
\end{eqnarray}

The high accuracies (SM, Fig. S3) of the above forms further verify the regime classification. Hence, a formal reformulation of PB theory is established, with each ($\gamma$,$\chi$) corresponding to one regime and specifying the distinct EDL feature. The regime-separation shown in Fig.~\ref{fig:Figure 1}(i) is the source of nanofluidic transport diversities.

\section{Field-tunable nanofluidic transport}
\begin{figure*}[htbp!!!!!!!]
	\includegraphics[width=1\textwidth]{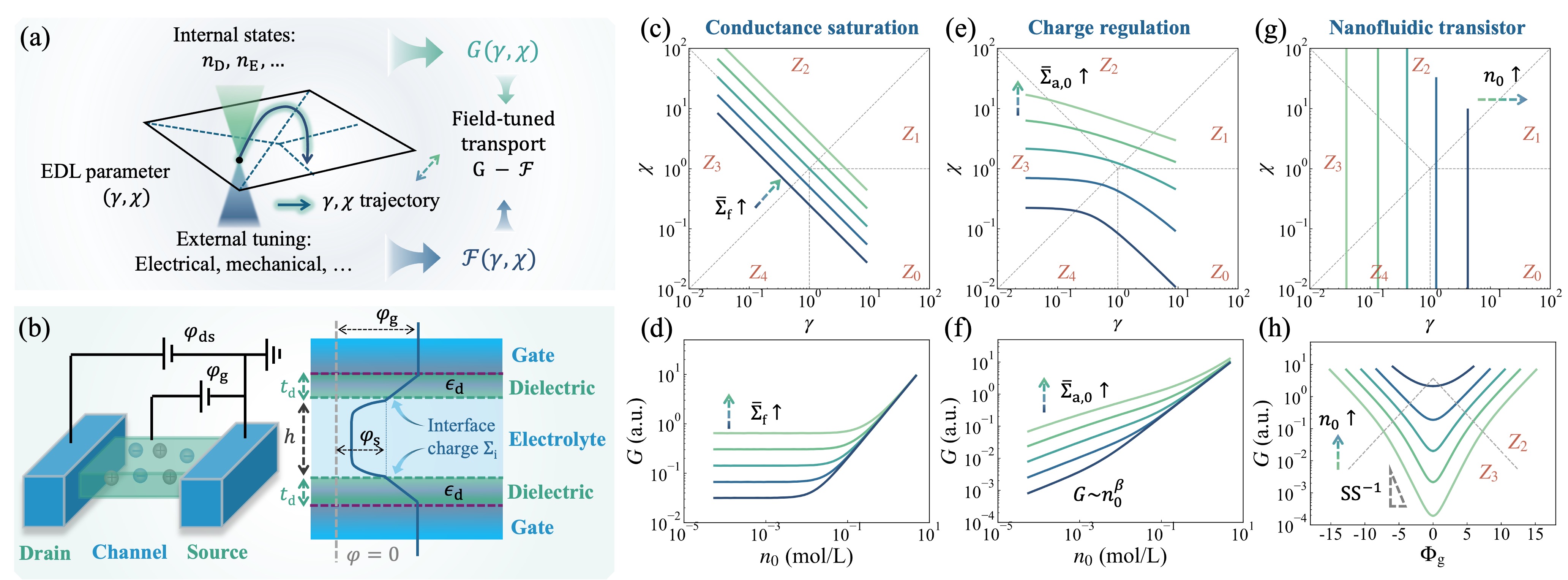}% Here is how to import EPS art
	\caption{(a) The framework for field-tuned transport $G - \mathcal{F}$. (b) Schematics of a nanofluidic device and the
		gate-dielectric-electrolyte structure. (c, d) Conductance
		saturation with  $\overline{\Sigma }_\mathrm{f}$ being 0.25, 0.5, 1, 2 and 4. (e, f) Charge regulation with  $\overline{\Sigma }_{\mathrm{a},0}$ being 0.1, 1, 10, 100 and 1000. (g, h) The ideal transistor at  $\eta _\mathrm{g}\rightarrow {\infty}$, with  $n_0$ being 0.1 mM, 1 mM, 10 mM, 0.1 M and 1M. For
		(d)(f)(h),  $h=5$ \ nm and  $l_\mathrm{B}=0.7$ \ nm.}
	\label{fig:Figure 2}
\end{figure*}

A universal and formal framework for field-tunable nanofluidic transport is built. Firstly, device parameters can be mapped to ($\gamma$,$\chi$), which determines the inner EDL states ($n_\mathrm{D},n_\mathrm{E}$) and the conductivity $G$. Secondly, the external tuning parameters $\mathcal{F}$  can be captured by the modulation of $\gamma$ and  $\chi$. Thus by establishing the respective linkages of $G$ and $\mathcal{F}$ to ($\gamma$,$\chi$), the field-tuned transport $G(\mathcal{F})$ can be resolved, as illustrated in Fig.~\ref{fig:Figure 2}(a).
\begin{equation}
	G(\gamma,\chi), \mathcal{F}(\gamma,\chi)\quad\rightarrow \quad G(\mathcal{F}).
	\label{eq:F_G_relation}
\end{equation}
Different types of $\mathcal{F}$ generate different trajectories on the ($\gamma$,$\chi$) plane and bring diverse transport behaviors $G(\mathcal{F})$. 

In this work, we focus on the electrostatic modulation via a gate voltage  $\Phi _\mathrm{g}$, which is pivotal to numerous iontronic applications \cite{017,036,043}. Consider the ionic conductivity $G(\gamma,\chi)$ in the simplest scenario (steady
state, constant mobility $\mu$):
\begin{equation}
	G=\mathit{e\mu }\left(2n_0+n_\mathrm{E}+n_\mathrm{D}\right)=2n_0\mathit{e\mu }\left(1+\Gamma
	\right).
	\label{eq_conductivity}
\end{equation} A gate-dielectric-electrolyte structure is considered, as shown in Fig.~\ref{fig:Figure 2}(b). The external gate voltage  $\Phi _\mathrm{g}$  couples to the EDL surface potential  $\Phi _\mathrm{s}$, through the electric field  $E_\mathrm{d}$ transmitted in the dielectric layer ($\epsilon_\mathrm{d}, t_\mathrm{d}$):
\begin{equation}
	\varphi _\mathrm{g}=\varphi _\mathrm{s}+E_\mathrm{d}t_\mathrm{d},\quad\epsilon E_\mathrm{s}=\epsilon _\mathrm{d}E_\mathrm{d}+e\Sigma _\mathrm{i}.
	\label{eq_potential_connection}
\end{equation}
The charge $\Sigma _\mathrm{i}$ at the electrolyte/dielectric interface can be categorized into two types: one with a fixed number density  $\Sigma _\mathrm{f}$ \cite{010}, and the other with a tunable density $\Sigma _\mathrm{a}$ by  $\varphi _\mathrm{s}$ \cite{044}. 
With Bjerrum length  $l_\mathrm{B}=e^2/4\mathit{\pi \epsilon k}_\mathrm{B}T$, $\Sigma _\mathrm{i,f,a}$ are reduced to dimensionless forms $\overline{\Sigma }_\mathrm{i,f,a}=\Sigma _\mathrm{i,f,a}\cdot\left(\frac{\pi } 2hl_\mathrm{B}\right)$ and the linkage of $\Phi _\mathrm{g}$ to $(\gamma,\chi)$ is established:
\begin{equation}
	\Phi _\mathrm{g}=\Phi _\mathrm{s}(\gamma,\chi)+\frac 1{\eta _\mathrm{g}}\left[\mathit{\gamma \chi }-\overline{\Sigma }_\mathrm{f}-\overline{\Sigma}_\mathrm{a}\left(\Phi _\mathrm{s}\right)\right],
	\label{eq_master_equation}
\end{equation}
where $\eta _\mathrm{g}=\epsilon _\mathrm{d}h/8\epsilon t_\mathrm{d}$ is the dielectric coupling efficiency. The gate-tuned transport $G-\Phi_\mathrm{g}$ can now be described. 

The formulation of eq.~(\ref{eq_master_equation}) can be validated by the faithful reproductions of conductance saturation and charge regulation phenomena. In the floating gate state ($\eta _\mathrm{g}\rightarrow0$, $\Phi_\mathrm{g}$ is irrelevant and $n_0$ can act as the tuning parameter), the transport is governed by the interfacial charges: $	\mathit{\gamma \chi }-\overline{\Sigma }_\mathrm{f}-\overline{\Sigma }_\mathrm{a}\left(\Phi
_\mathrm{s}\right)=0.$
For  $\overline{\Sigma }_\mathrm{a}=0$,  $\mathit{\gamma \chi }=\overline{\Sigma }_\mathrm{f}$  yields straight trajectories in the  $(\gamma ,\chi )$ plane (Fig.~\ref{fig:Figure 2}(c)). And as shown in Fig.~\ref{fig:Figure 2}(d), $G$ saturates at small $n_0$, as have been observed in early studies \cite{010,011}. As for $\overline{\Sigma }_\mathrm{f}=0$,  $\mathit{\gamma \chi
}=\overline{\Sigma }_\mathrm{a}(\Phi _\mathrm{s})$ \ corresponds to the charge regulation effect \cite{044,045}, where interfacial charges
adjust to  $\Phi _\mathrm{s}$, usually in the form of  $\overline{\Sigma }_\mathrm{a}=\overline{\Sigma }_{\mathrm{a},0}\exp (\pm \Phi_\mathrm{s})$. Curved trajectories are yielded in the $(\gamma ,\chi )$ plane (Fig.~\ref{fig:Figure 2}(e)) along with an unconventional scaling behavior for conductance in the form of $G\propto n_0^{\beta }$ (Fig.~\ref{fig:Figure 2}(f)). This work finds  $\beta$ ranging from 0.4 to 0.6 depending on $\overline{\Sigma }_{\mathrm{a},0}$ (SM,
Fig. S4), which is close to the observed values ranging from 1/3 to 2/3 \cite{013,014,020,046}.  And more significantly, the theoretically predicted $G(n_0)$ relation accurately aligns with the experimental data measured on our SiO$_2$ nanochannel (SM, Fig. S7), and the value of $\overline{\Sigma }_{\mathrm{a},0}$ can be identified via a nonlinear fitting process. These results not only validate the proposed theoretical framework, but also open a plausible route to evaluate, predict and optimize realistic field-tunable devices, which is shown next.
\begin{figure}[htbp!!!!!]
	\includegraphics[width=0.48\textwidth]{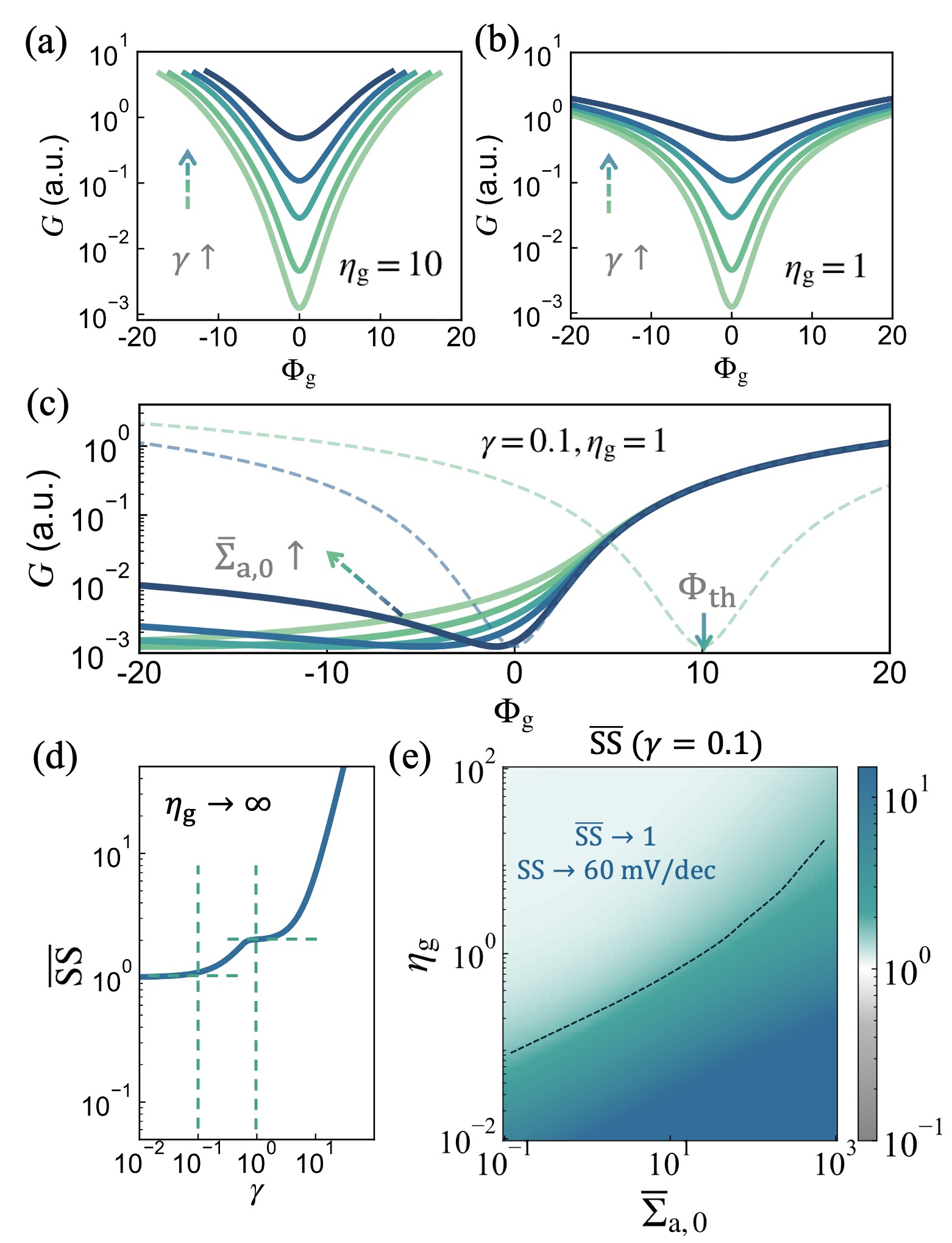}% Here is how to import EPS art
	\caption{Transistor behaviors at realistic conditions. (a, b) Transfer characteristic curves at $\eta _\mathrm{g}=10$ and 
		$\eta _\mathrm{g}=1$, with $\gamma$ being 0.1, 0.2, 0.5, 1 and 2 (c) Transfer characteristic curves with non-zero
		$\overline{\Sigma }_\mathrm{f}$ \ and $\overline{\Sigma}_{\mathrm{a},0}$.  $\overline{\Sigma}_{\mathrm{a},0}$ takes 0, 1, 5, 10,
		20 and 40. (d) Ideal $\overline{\mathrm{SS}}$ as a function of $\gamma$. (e) Realistic $\overline{\mathrm{SS}}$ at $\gamma=0.1$.}
	\label{fig:Figure 3}
\end{figure}

Now start the discussion of $G-\Phi_\mathrm{g}$ relations (ionic transistor behavior) with an ideal scenario $\eta _\mathrm{g}\rightarrow {\infty}$ (equivalently  $\Phi _\mathrm{s}=\Phi _\mathrm{g}$). By tracing vertical lines as shown in Fig.~\ref{fig:Figure 2}(g), ambipolar transfer characteristic curves at different  $n_0$ ($\gamma$) are plotted in Fig.~\ref{fig:Figure 2}(h). Upon applying a positive (negative)  $\Phi _\mathrm{g}$, anions
(cations) are enriched in the channel, leading to an exponential increase in conductivity. For large $n_0$ ($\gamma$), EDLs occupy only a small fraction of the channel, making the tunability and the on-off ratio (OOR,  OOR $=G_{\mathrm{max}}/G_{\mathrm{min}}$) less pronounced. In contrast, for a small  $n_0$ ($\gamma\ll1$), significant EDL-overlap induces a high tunability efficiency and an on-off ratio up to 10\textsuperscript{4}. Hence, despite the lack of quantum-mechanical band alignment effects as in electronic transistors, satisfying ionic switching performances can still be obtained via nanoconfinement. Next consider finite $\eta _\mathrm{g}$ and adopt subthreshold swing (SS) from electronic transistors, defined as $\mathrm{SS}=\mathrm{d}\varphi _\mathrm{g}/\mathrm{d}(\log G)$. Due to the dielectric coupling, a portion of the applied  $\Phi_\mathrm{g}$ drops across the dielectric and is ineffective for modulation. Hence, from Fig.~\ref{fig:Figure 3}(a) to Fig.~\ref{fig:Figure 3}(b), as  $\eta_\mathrm{g}$ decreases from 10 to 1, a larger  $\Phi _\mathrm{g}$ is required to achieve the same conductivity (or equivalently, the same  $\chi$) as in Fig.~\ref{fig:Figure 3}(a), leading to an  increase in $\mathrm{SS}$. Finally, consider the most realistic scenario where  $\eta _\mathrm{g}$ is finite and interfacial charges are involved. In Fig.~\ref{fig:Figure 3}(c), the original transfer curve (dashed blue line) is shifted (to the dashed green line) due to a
non-zero $\overline{\Sigma }_\mathrm{f}$, indicating that a fixed interfacial charge density introduces a threshold-voltage ($\Phi_	\mathrm{th}$) shift. As shown by the solid lines in Fig.~\ref{fig:Figure 3}(c), the charge-regulation effect $\overline{\Sigma }_\mathrm{a}=\overline{\Sigma }_{\mathrm{a},0}\exp (-\Phi _\mathrm{s})$ \ dramatically converts the ambipolarity to a
single N-polarity. As  $\overline{\Sigma }_{\mathrm{a},0}$ increases, the tunability is gradually weakened. The disappearance of P-polarity at negative gate potential in Fig.~\ref{fig:Figure 3}(c) arises from the accumulation of positive interfacial charges, which increase exponentially with  $|\Phi _\mathrm{s}|$, offset the external electric field  $E_\mathrm{d}$ and in turn constrain $\chi$. Complementarily, a single P-polarity can in principle be achieved by another form of  $\overline{\Sigma }_\mathrm{a}=-|\overline{\Sigma }_{\mathrm{a},0}| \exp (\Phi _\mathrm{s})$. These results prove that the polarities of ionic
transistors can be effectively reconfigured via surface modifications, making them promising candidates for new information processing paradigms. 
\begin{table}
	\centering
		\caption{Summary of ion transport.}
	\label{tab:Table_2}
	\renewcommand{\arraystretch}{1.35}
	\setlength{\tabcolsep}{3.0pt}
	\begin{tabular}{|c|c|c|c|c|}
		\hline
		$\eta_\mathrm{g}$ & $\overline{\Sigma }_{\mathrm{f}}$ & $\overline{\Sigma }_{\mathrm{a},0}$ & \begin{tabular}{c}
			$\mathcal{F}$
		\end{tabular}& \begin{tabular}{c}
			$G(\mathcal{F})$
		\end{tabular}  \\
		\hline
		0&Finite&0&$n_0$&\begin{tabular}{c}
			$G$ saturation
		\end{tabular}\\
		\hline
		0&/&Finite&$n_0$&\begin{tabular}{c}
			$G\propto n_0^{\beta}$
		\end{tabular}\\
		\hline
		$\infty$& / & / &$\Phi_\mathrm{g}$& \begin{tabular}{c}
			Ideal ambipolar transistor
		\end{tabular}\\
		\hline
		Finite&Finite&0&$\Phi_\mathrm{g}$&\begin{tabular}{c}
			Transistor with $\Phi_\mathrm{th}\neq0$
		\end{tabular}\\
		\hline
		Finite&Finite&Finite&$\Phi_\mathrm{g}$&\begin{tabular}{c}
			Transistor of	single polarity		\end{tabular}\\
		\hline
	\end{tabular}
\end{table}

Quantitative investigations upon $\mathrm{SS}$ and on-off ratio are conducted, as they are key performance indicators for ionic transistors. The ideal $\eta_\mathrm{g}\rightarrow\infty$ produces the smallest  $\overline{\mathrm{SS}}$ ( $\overline{\mathrm{SS}}=\mathrm{d}\Phi _\mathrm{g}/\mathrm{d}(\ln G)$,  $\mathrm{SS}=\ln
10{\cdot}\overline{\mathrm{SS}}{\cdot}k_\mathrm{B}T/e$) shown in Fig.~\ref{fig:Figure 3}(d). Under prominent confinement ($\gamma <0.1$), the $\overline{\mathrm{SS}}$ approaches 1 but no smaller. As  $\gamma$ increases from 0.1 to 1, $\overline{\mathrm{SS}}$ increases to 2 and saturates. And when  $\gamma$ reaches 10, $\overline{\mathrm{SS}}$ loses its practical meaning as the ion transport is bulk-like. Under realistic conditions, $\overline{\mathrm{SS}}$ increases, which can be
estimated by eq.~(\ref{eq_SS_apporx}) and is shown in Fig.~\ref{fig:Figure 3}(e) for $\gamma =0.1$.
\begin{equation}
	\overline{\mathrm{SS}}=\overline{\mathrm{SS}}\left(\gamma, \eta _\mathrm{g}\rightarrow {\infty}\right)+\frac{2\gamma^2+|\overline{\Sigma}_{\mathrm{a},0}|}{2\eta_\mathrm{g}}.
	\label{eq_SS_apporx}
\end{equation}
$\overline{\mathrm{SS}}\rightarrow 1$ and  $\overline{\mathrm{SS}}\rightarrow 2$ are thus two significant limits. $\overline{\mathrm{SS}}\rightarrow 1$ occurs in the  $n_\mathrm{D}$ regime where $\varphi(z)$ is almost uniform across the channel, yielding $k_\mathrm{B}T\ln n_\mathrm{D}/n_0-e\varphi _\mathrm{s}=0$ from the minimization of
free energy. Therefore $n_\mathrm{D}\propto\exp (\Phi _\mathrm{s})$ and  $\overline{\mathrm{SS}}\rightarrow 1$. $\overline{\mathrm{SS}}\rightarrow 2$ occurs in the  $n_\mathrm{E}$ regime where the potential quickly decays to zero and the effective potential imposed to the EDL is the average value  $\varphi _\mathrm{s}/2$. Hence  $n_\mathrm{E}\propto\exp (\Phi _\mathrm{s}/2)$ and $\overline{\mathrm{SS}}\rightarrow 2$. The two limits have values of 60~mV/dec and 120~mV/dec at 300 K, with the former one echoing thermodynamic limit in electronic transistors. For on-off ratio, it can be concluded from Figs.~\ref{fig:Figure 3}(b) and (c) that the finite $\eta _\mathrm{g}$ constrains the conductance from further increasing in the high-$\Phi_\mathrm{g}$ regime, and that a nonzero $\overline{\Sigma}_{\mathrm{a},0}$ interrupts the electric neutrality and lifts the conductance in the $\Phi_\mathrm{g} \rightarrow 0$ regime. Since the applied $\Phi_\mathrm{g}$ cannot be unlimitedly large, the practical OOR under working conditions is predicted by:
\begin{equation}
	\mathrm{OOR}=\frac{\eta_{\mathrm{g}}}{\gamma^2}\cdot\frac{10}{\sqrt{1+(\frac{\overline{\Sigma}_{\mathrm{a},0}}{2\eta_{\mathrm{g}}})^2}}
	\label{eq_OOR_apporx}
\end{equation}
\begin{figure}[htbp]
	\includegraphics[width=0.48\textwidth]{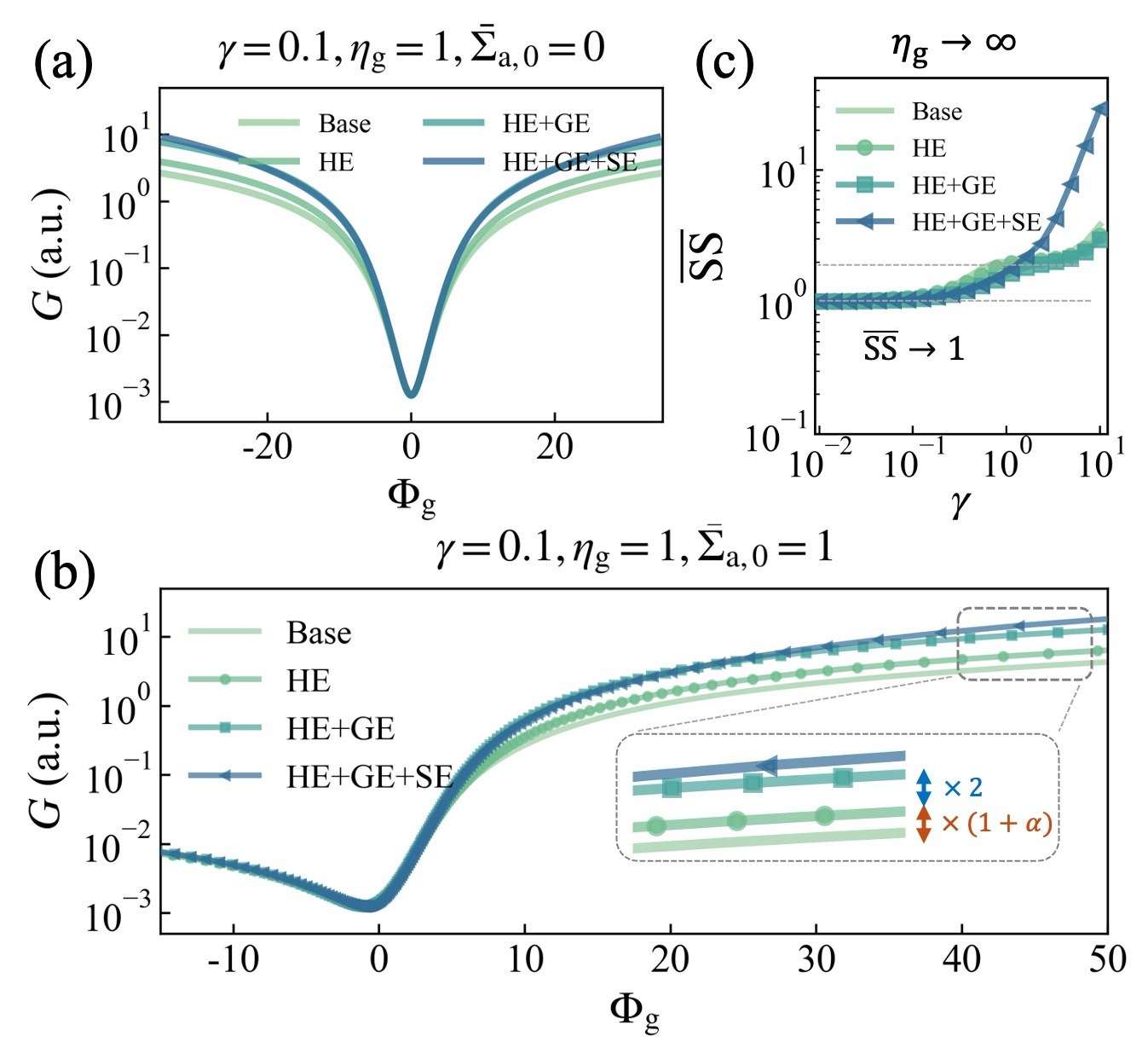}% Here is how to import EPS art
	\caption{Transistor behaviors considering more sophisticated effects. (a) Transfer characteristic curves at $\gamma=0.1, \eta_{\mathrm{g}}=1$ and $\Sigma_{\mathrm{a},0}=0$. (b) Transfer characteristic curves at $\gamma=0.1, \eta_{\mathrm{g}}=1$ and $\Sigma_{\mathrm{a},0}=1$. (c)  Ideal $\overline{\mathrm{SS}}$ as a function of $\gamma$ after considering HE, GE,and SE.}
	\label{fig:Figure 4}
\end{figure}

Above a comprehensive analysis over electrostatic modulation has been done upon the simplest form of conductance in eq. (\ref{eq_conductivity}). Upon this minimal setting, the principal physical mechanisms together with a set of key device parameters ($\gamma$, $\overline{\Sigma}_{\mathrm{a},0}$ and $\eta_{\mathrm{g}}$) have been identified and summarized in Table.~\ref{tab:Table_2}. We now further incorporate more sophisticated yet equally realistic effects, including hydrodynamic effect (HE) and geometric effect (GE). The analytical derivations are provided in SM, section 2 and we plot the transfer characteristic curves in Figs. \ref{fig:Figure 4} (a) and (b). In the high-$\Phi_\mathrm{g}$ regime (also the $n_\mathrm{E}$ regime), the hydrodynamic effect (electro-osmosis) increases the conductance by a factor of $1+\alpha$, with $\alpha=2\epsilon k_\mathrm{B}T/e\mu\eta \approx 1/2$, where $\eta$ is the viscosity. Changing the nanochannel from the nanoslit configuration to the nanotube configuration, the geometric effect increases the conductance further by 2, which can be intuitively explained by the portion of EDLs (for thin EDL layers with $l_\mathrm{D}$, they take up $2l_\mathrm{D}/h$ of the nanoslit and $4l_\mathrm{D}/h$ of the nanotube). However, neither of the two effects alters the scaling behavior of conductance in the sub-threshold regime (also the $n_\mathrm{D}$ regime) and the ideal SS values shown in Fig. \ref{fig:Figure 4} (c) are almost identical to those shown in Fig. \ref{fig:Figure 3} (d). Finally, we consider the size effect (SE) formulated by $n_{\pm }(z)=n_0\exp (-z_{\pm }\Phi)/[1-\eta_0+\eta_0 \cosh(z_{\pm }\Phi)]$ \cite{032}, where $\eta_0$ is the volume factor. The size effect barely alters the conductance in the high-$\Phi_\mathrm{g}$ regime but it does influence the SS values. When $\gamma>1$, SS does not saturate at 2 but increases all along (Fig. \ref{fig:Figure 4} (c)). As ions have finite sizes, the accumulation layer cannot be unrealistically compressed, especially when $l_\mathrm{GC}$ approaches the ion diameter. Hence the continuum description in the  $n_\mathrm{E}$ regime may break down and thus $\overline{\mathrm{SS}}\rightarrow 2$ may be invalid. Nevertheless,  $\overline{\mathrm{SS}}\rightarrow 1$ is robust even after considering these sophisticated effects, and is therefore a thermodynamic limit. Since the ionic enrichments in EDLs couple to mechanical, hydrodynamic, and interfacial degrees of freedom, the SS limit is expected to as well bound the tunability of other EDL-mediated properties. In this sense, the SS limit identified here constitutes an universal and fundamental constraint on the modulation of nanofluidic properties.

\section{Conclusions and outlooks}
Based on the regime classification of the confined-EDL problem, a formal framework for field-tunable nanofluidic transport is established. By linking ionic conductivity $G$ to voltage $\Phi_\mathrm{g}$ via EDL parameters ($\gamma$,$\chi$), ionic transistor behaviors are accurately and extensively studied, and the fundamental thermodynamic limit is discovered for electrostatic modulation. 

These findings deepen the understanding of ion transport under nanoconfinement and provide direct guidance for experimental optimization of nanofluidic devices. Using eqs. (\ref{eq_SS_apporx}) and (\ref{eq_OOR_apporx}), and selecting appropriate device parameters ($\gamma$, $\overline{\Sigma}_{\mathrm{a},0}$ and $\eta_{\mathrm{g}}$), the field-tunable devices with expected performances (SS, OOR and polarity) can be designed in prior. Moreover, these results can serve as a mean-field benchmark, against which beyond-mean-field phenomena can be confirmed, when entering the sub-1 nm realm \cite{049,050,051,052}.

Starting with the minimal setting in eq. (\ref{eq_conductivity}), we successfully extend the framework with more sophisticated physical effects, including hydrodynamic enhancement, geometric curvature enhancement and size effect in the $n_\mathrm{E}$ regime. Furthermore, this framework $G(\mathcal{F})$ can naturally be generalized to consider other types of stimuli (e.g., mechanical \cite{048}). Hence, being accurate, extensible and generalizable, this work shall apply to a wide scope.

\begin{acknowledgments}
This work was financially supported by the National Key Research and Development Program of China (Grant No.
2022YFA1203400), the National Natural Science Foundation of China (Grant No. 61774090) and the Key-Area Research and
Development Program of Guangdong Province (Grant No. 2020B010169001).

\end{acknowledgments}
\bibliography{manuscript_v8}% Produces the bibliography via BibTeX.

\end{document}